\documentclass[reprint,aps,superscriptaddress,nofootinbib,preprintnumbers]{revtex4-2}

\usepackage{amsmath,amsthm,amsfonts}
\usepackage{graphicx}
\usepackage{physics}
\usepackage{multirow}
\usepackage{hyperref}
\hypersetup{
	colorlinks   = true, 
	urlcolor     = blue, 
	linkcolor    = blue, 
	citecolor    = red   
}

\graphicspath{{figs/}}

\newcommand{\nc}{\newcommand}

\nc{\calR}{{\cal{R}}}
\nc{\calP}{{\cal{P}}}
\nc{\cN}{ {\cal{N}} }

\usepackage{tikz}
\usetikzlibrary{arrows,shapes}
\usetikzlibrary{trees}
\usetikzlibrary{matrix,arrows} 				
\usetikzlibrary{positioning}				
\usetikzlibrary{calc,through}				
\usetikzlibrary{decorations.pathreplacing}  
\usepackage{pgffor}							

\begin{document}


\title{k-inflation: Non-separable case meets ACT measurements}

\author{Tahere Fallahi Serish}
\email{fallahi.ta@shahroodut.ac.ir}
\affiliation{Faculty of Physics, Shahrood University of Technology, P.O. Box 3619995161 Shahrood, Iran}

\author{Seyed Ali Hosseini Mansoori}
\email{shosseini@shahroodut.ac.ir}
\affiliation{Faculty of Physics, Shahrood University of Technology, P.O. Box 3619995161 Shahrood, Iran}

\author{Fereshteh Felegary}
\email{fereshteh.felegary@gmail.com}
\affiliation{School of Astronomy, Institute for Research in Fundamental Sciences (IPM), Tehran, Iran, P.O. Box 19395-5531}

\author{\"{O}zg\"{u}r Akarsu}
\email{akarsuo@itu.edu.tr}
\affiliation{Department of Physics, Istanbul Technical University, Maslak 34469 Istanbul, T\" urkiye}

\author{M. Sami}
\email{sami\_ccsp@sgtuniversity.org}
\affiliation{Centre for Cosmology and Science Popularization (CCSP), SGT University, Gurugram, Delhi- NCR, Haryana- 122505, India}
\affiliation{Eurasian International Centre for Theoretical Physics, Astana, Kazakhstan}


\begin{abstract}
We investigate a non-separable subset of $k$-essence in which the kinetic and potential sectors interact through an $X^{\rho}V(\phi)$ coupling, implemented via a potential-dependent prefactor $f(\phi)=1+2\mathcal{K}V$. In slow roll, this structure preserves a constant sound speed $c_s^2=1/(2\rho-1)$ while modifying the Hubble flow in a controlled way, thereby shifting the inflationary observables relative to the separable template. For monomial potentials $V=A\phi^n$ (with $n=2$ and $n=2/3$ as representative cases) we derive closed analytic expressions for $n_s(N_\ast)$ and $r(N_\ast)$ to $\mathcal{O}(\epsilon_{\rm mix}^2)$, where $\epsilon_{\rm mix}\propto\mathcal{K}$ encodes the non-separable $X^\rho V$ mixing, and we validate them against exact background integrations. The analytic and numerical predictions agree at the sub-per-mille level for $n_s$ and at the percent level for $r$, confirming the accuracy of the small-mixing expansion. For $\mathcal{K}<0$ the mixing systematically lowers both $n_s$ and $r$ at fixed $N_\ast$, allowing otherwise marginal monomials to fall within the region favored by recent ACT+{\it Planck}+BAO constraints (P--ACT--LB). All solutions shown satisfy the health conditions $f(\phi)>0$, $\rho>\tfrac12$, and the positivity bound $V<1/(2|\mathcal K|)$ (from $f>0$). We also discuss parameter dependence and the expected equilateral-type non-Gaussianity, which remains comfortably within current bounds for the benchmarks considered.
\end{abstract}

\maketitle 


\section{Introduction}\label{sec0}

Inflation provides a compelling mechanism to resolve the horizon, flatness, and monopole problems of standard big–bang cosmology while generating the primordial fluctuations that seed large–scale structure~\cite{Guth1981Inflation,Linde1982PBL,Starobinsky1980PBL,MukhanovChibisov1981JETPL,AlbrechtSteinhardt1982PRL}. Quantum fluctuations of the inflaton field source metric perturbations, whose imprints on the cosmic microwave background (CMB) are captured, to leading order, by two key observables: the scalar spectral index $n_s$, encoding the scale dependence of curvature perturbations, and the tensor–to–scalar ratio $r$, measuring the relative amplitude of primordial gravitational waves.

Over the past decades, successive space and ground experiments—from COBE and WMAP~\cite{Hinshaw2013WMAP} to Planck~\cite{Planck:2018jri}, ACT~\cite{ACT:2025fju,ACT:2025tim}, and SPT~\cite{Balkenhol2023SPT3G}—have progressively sharpened these constraints. The Planck–only value $n_s=0.9649\pm0.0042$~\cite{Planck:2018jri} is shifted upward in joint ACT+Planck analyses (P--ACT), which find $n_s=0.9709\pm0.0038$; including BAO/DESI information (P--ACT--LB) further yields $n_s=0.9743\pm0.0034$ together with $r<0.038$ (95\% CL)~\cite{ACT:2025fju,ACT:2025tim}. For comparison, the current SPT/Planck combination gives $n_s=0.9647\pm0.0037$~\cite{Balkenhol2023SPT3G}. The upward shift in $n_s$ is phenomenologically significant: it places pressure on a number of otherwise successful benchmarks and has motivated a broad set of refinements and alternatives, including deformations of Starobinsky/attractor realizations, non/minimally coupled models, and higher–curvature frameworks~\cite{Ellis:2025ieh,Antoniadis:2025arXiv,Gialamas:2025arXiv,Addazi:2025arXiv,Mohammadi:2025arXiv,Haque:2025arXiva,Haque:2025arXivb,Drees:2025arXiv,Kallosh:2025arXiv,Aoki:2025arXiv,Mondal:2025arXiv,Liu:2025arXiv,Maity:2025arXiv,Gao:2025arXiv,Dioguardi:2025arXiv,McDonald:2025arXiv,Yin:2025arXiv,Yi:2025arXiv,Peng:2025arXiv,He:2025arXiv,Wolf:2025arXiv,Berera:2025arXiv,Pallis:2025nrv,Odintsov:2025bmp,Zhu:2025twm,Ketov:2025cqg,Zahoor:2025nuq,Chakraborty2025_BehaviourAlphaAttractors_WarmInflation,Han2025_HiggsPoleInflation_ACT,Gao2025_ObservationalConstraints_NMC_ACT,Choudhury2025_NewPhysics_ACT_DESI,Gao2025_ObservationalConstraints_NMC_ACT,Yi2025_PolynomialAlphaAttractor,He2025_IncreaseNS_PoleInflation_EC,Heidarian:2025drk,Oikonomou:2025xms,Yuennan:2025kde,Oikonomou:2025htz,Odintsov:2025jky}.

Within single–field scenarios, $k$-inflation~\cite{Armendariz-Picon:1999hyi} (or, more broadly, $k$–essence) retains minimal coupling to gravity while generalizing the Lagrangian to $P(X,\phi)$, with $X\equiv -\partial_\mu\phi\,\partial^\mu\phi/2$. This framework admits a subluminal sound speed $c_s\le1$ and, in general, distinctive non–Gaussian signatures. Most phenomenological applications adopt a sum–separable structure, $P(X)+U(\phi)$. In this work we explore a simple non-separable template where the kinetic and potential sectors interact multiplicatively via an $X^\rho V(\phi)$ coupling, encoded by
$P(X,\phi)=f(\phi)\,[X^{\rho}-V(\phi)]$ with $f(\phi)=1+\frac{2\mathcal K}{M_{\rm pl}^4}V(\phi)$ and $\rho>\tfrac12$. This structure preserves a constant sound speed $c_s^2=1/(2\rho-1)$ and modifies the slow-roll Hubble flow in a controlled way, thereby shifting $(n_s,r)$ relative to the separable template. The sign of the mixing parameter $\mathcal K$ is model–dependent; here we focus on $\mathcal K<0$, for which $0<f(\phi)<1$ during slow roll and the mixing remains perturbative, while predictions tend to move toward the recent P--ACT--LB preferred region. (An EFT rationale for $f(\phi)$ and its sign is given in Sec.~\ref{Sec1}.)

Our goals are twofold. First, for monomial potentials $V=A\phi^n$ (with $n=2$ and $n=2/3$ as representative cases), we derive closed analytic expressions for $n_s(N_\ast)$ and $r(N_\ast)$ to $\mathcal O(\epsilon_{\rm mix}^2)$, where the small parameter $\epsilon_{\rm mix}\propto\mathcal K$ naturally encodes the non–separable $X^\rho V$ mixing. We validate these expressions against exact background integrations, finding excellent agreement in the small–mixing regime. Second, we map the dependence of $(n_s,r)$ on $(N_\ast,\rho,\epsilon_{\rm mix})$ and compare the resulting predictions with recent ACT+Planck+BAO contours (P--ACT--LB) via parameter–space overlays in the $(n_s,r)$ plane. This investigation clarifies the roles of $N_\ast$ (primarily setting $n_s$), $\rho$ (via the constant $c_s$), and the mixing strength $\epsilon_{\rm mix}$ (which tunes both observables), and it highlights regimes in which otherwise marginal monomials become compatible with current bounds. In particular, we show that even for simple chaotic monomials the non–separable mixing can move predictions into the P--ACT--LB–favored region.

The paper is organized as follows. In Sec.~\ref{Sec1} we introduce the setup and slow-roll background, including stability conditions and the constant sound speed. In Sec.~\ref{Sec2} we derive analytic formulae for the slow-roll parameters and observables in terms of $(N_\ast,\rho,\epsilon_{\rm mix})$. In Sec.~\ref{Sec3} we compare these expressions with exact integrations and recent constraints, presenting representative tables and parameter–space plots. We summarize and discuss future directions in Sec.~\ref{Sec4}.

\section{Setup}\label{Sec1}

Among $k$-essence models~\cite{Chen_2007,Armendariz-Picon:1999hyi}, we are led to consider a simple but instructive non-separable template in which the kinetic and potential sectors do not merely add but interact multiplicatively. Concretely, we take
\begin{equation}\label{Lmnon}
P(X,\phi)=f(\phi)\big[X^{\rho}-V(\phi)\big],
\end{equation}
where $X \equiv -\tfrac12\,\partial_{\mu}\phi\,\partial^{\mu}\phi$ is the canonical kinetic invariant, $V(\phi)$ is the inflaton potential, and $\rho>0$ is a constant exponent controlling the departure from canonical dynamics. The hallmark of~\eqref{Lmnon} is the prefactor $f(\phi)$, which we choose to depend on the potential itself,
\begin{equation}\label{f}
f(\phi)=1+\frac{2\mathcal{K}}{M_{\rm pl}^{4}}\,V(\phi),
\end{equation}
with $\mathcal{K}$ a dimensionless coupling. We henceforth work in reduced Planck units, $M_{\rm pl}=1$, so $f(\phi)=1+2\mathcal K V$. This choice captures the lowest-order non-separable mixing between $X$ and $V$ in a low-energy expansion: the Lagrangian effectively contains an $X^{\rho}V$ interaction whose strength is modulated by $f(\phi)$. In the limit $\mathcal{K}\to0$, the mixing switches off and~\eqref{Lmnon} reduces continuously to the familiar separable, non-canonical case $P(X)+U(\phi)$~\cite{Li_2012}. We work in reduced Planck units ($M_{\rm pl}=1$) unless stated otherwise.

The potential dependence of $f$ admits a natural rationale in effective field theory (EFT).\footnote{One convenient route is to start from a multi-field completion where heavy modes couple to both the inflaton potential and its kinetic sector. After integrating out the heavy degrees of freedom, the single-clock description can acquire a non-separable $P(X,\phi)$ with $\phi$-dependent Wilson coefficients. Expanding those coefficients in $V/M_{\rm pl}^{4}$ (or, equivalently, in inverse powers of the heavy mass scale) yields $f(\phi)=1+c_1\,V/M_{\rm pl}^{4}+\cdots$. Retaining the lowest-order correction and identifying $c_1\equiv 2\mathcal{K}$ reproduces~\eqref{f} and, with~\eqref{Lmnon}, generates the lowest-order $X^{\rho}V$ mixing beyond the separable template. See, e.g., Refs.~\cite{Cheung:2007st,Achucarro:2010jv,Achucarro:2012hg}.} The sign of $\mathcal{K}$ is not dictated \textit{a priori} by this EFT reasoning—both signs can arise depending on the UV couplings. In what follows we focus on the regime $\mathcal{K}<0$. For positive $V(\phi)$ during slow roll, this choice keeps the prefactor in the controlled window $0<f(\phi)<1$ and implies the bound $V(\phi)<M_{\rm pl}^{4}/(2|\mathcal{K}|)$, which in our units reads $V<1/(2|\mathcal K|)$. This ensures that the non-separable mixing remains perturbative, $|\mathcal K|V\ll1$. Phenomenologically, this sign typically shifts the inflationary observables $(n_s,r)$ downward relative to the separable limit for the potentials we study, improving agreement with the ACT-preferred region; quantitative comparisons are deferred to the Results.

In summary, our working hypothesis~\eqref{Lmnon}–\eqref{f} provides a minimal arena in which a potential-dependent prefactor modulates non-canonical dynamics. The multiplicative structure is the key novelty—turning the sum of two sectors into an interaction—and the special case $\rho=1$ smoothly connects to the $XV$ model analyzed in~\cite{HosseiniMansoori:2024pdq}. In the next section we derive the background equations implied by~\eqref{Lmnon} and track how the coupling in~\eqref{f} imprints on the slow-roll evolution and, ultimately, on the spectral index $n_s$ and the tensor-to-scalar ratio $r$.

\section{Background equations and Inflationary parameters}
\label{Sec2}

Working in a spatially flat FLRW spacetime and reduced Planck units ($M_{\rm pl}=1$), the background dynamics implied by~\eqref{Lmnon} are governed by the Friedmann relation
\begin{equation}\label{Hnonold}
3H^{2}=f\big[V+(2\rho-1)X^{\rho}\big]\,,
\end{equation}
and by the $k$-essence continuity (or field) equation, which can be written as
\begin{equation}\label{rhoback}
\begin{aligned}
6\rho\,f\,H\,X^{\rho}+\rho(2\rho-1)\,f\,X^{\rho-1}\dot X\\
= -\,\dot\phi\Big[(V+(2\rho-1)X^{\rho})\,f_{,\phi}+f\,V_{,\phi}\Big]\,,
\end{aligned}
\end{equation}
where a comma and a dot denote derivatives with respect to $\phi$ and $t$, respectively, and $f_{,\phi}=2\mathcal K\,V_{,\phi}$.

In the slow-roll regime we assume a potential-dominated energy budget and slowly varying kinetic energy,
\begin{equation}
(2\rho-1)\,X^{\rho}\ll V\,,\qquad \frac{|\dot X|}{HX}\ll1\,,
\end{equation}
under which the background equations simplify to
\begin{equation}\label{Hnon}
3H^{2}\simeq f\,V\,,
\end{equation}
\begin{equation}\label{vpnon}
6\rho\,H\,f\,X^{\rho}\simeq -\,\dot\phi\,V_{,\phi}\,(2f-1)\,,
\end{equation}
with $(2f-1)=1+4\mathcal K V$. Since $V>0$ during slow roll, Eq.~\eqref{Hnon} implies $f>0$ along the trajectory.

Finally, stability of cosmological scalar perturbations requires $P_{,X}>0$ and $c_s^2>0$~\cite{Chen_2007}. For the Lagrangian~\eqref{Lmnon}–\eqref{f} one finds
\begin{equation*}
P_{,X}=f\,\rho\,X^{\rho-1}\,,\qquad
c_s^2=\frac{P_{,X}}{P_{,X}+2X P_{,XX}}=\frac{1}{2\rho-1}\,,
\end{equation*}
so the conditions reduce to $f(\phi)>0$ and $\rho>\tfrac12$, with constant sound speed ($s\equiv\dot c_s/(Hc_s)=0$). In what follows we focus on $\mathcal K<0$; together with $V>0$ this implies the useful bound $V(\phi)<1/(2|\mathcal K|)$, which we impose in all parameter scans.

Having established the slow-roll background, it is convenient to relate the Hubble slow-roll parameter directly to potential data. Differentiating Eq.~\eqref{Hnon} gives
\begin{equation}\label{eq:epsHid}
\varepsilon_H \equiv -\frac{\dot H}{H^{2}}
\simeq -\frac{1}{2}\left(\frac{V_{,\phi}}{V}\right)
\left(\frac{\dot\phi}{H}\right)\left(\frac{2f-1}{f}\right).
\end{equation}
Next, using the slow-roll field relation~\eqref{vpnon} together with $X=\dot\phi^{2}/2$, one solves for the velocity–Hubble ratio as
\begin{equation}\label{phidoth1}
\frac{\dot\phi}{H}
= -\,C_\rho\left[
\frac{(2f-1)\,V_{,\phi}}{f^{\rho+1}\,V^{\rho}}
\right]^{\!\frac{1}{2\rho-1}},
\end{equation}
where $C_\rho \equiv \left(\frac{6^{\,\rho-1}}{\rho}\right)^{\!\frac{1}{2\rho-1}}$. For $\rho=1$ this reduces to
$\dot\phi/H = -\,(2f-1)\,V_{,\phi}/(f^{2}V)$, in agreement with~\eqref{vpnon} and $3H^{2}\simeq fV$. Combining~\eqref{eq:epsHid} with~\eqref{phidoth1} yields a compact expression for $\varepsilon_H$,
\begin{equation}\label{epsilonv}
\varepsilon_H \simeq \frac{C_\rho}{2}
\left[
\frac{(2f-1)^{2\rho}\,V_{,\phi}^{2\rho}}{f^{3\rho}\,V^{3\rho-1}}
\right]^{\!\frac{1}{2\rho-1}},
\end{equation}
while the exact relation in this model is $\varepsilon_H=\frac{3\rho\,X^{\rho}}{V+(2\rho-1)X^{\rho}}$, which reduces to $\varepsilon_H\simeq 3\rho X^{\rho}/V$ under the slow-roll conditions stated above.

During inflation the Hubble slow-roll parameters must remain small, 
$\varepsilon_H\ll1$ and $\eta_H\equiv \dot{\varepsilon}_H/(H\,\varepsilon_H)\ll1$, 
for roughly $50$–$60$ $e$-folds to solve the flatness and horizon problems. 
Inflation ends when $\varepsilon_H$ approaches unity; more generally, it is 
sufficient that either $\varepsilon_H$ or $|\eta_H|$ becomes $\mathcal{O}(1)$.

Additionally, the number of $e$-folds, $N$, measures the total inflation elapsed by its end (from a given field value to the end of inflation) and is
\begin{equation}\label{e-fold}
N=\int_{t}^{t_e}H\,dt
=\int_{V}^{V_e}\frac{1}{2\,\varepsilon_H}\,\frac{2f-1}{f}\,\frac{dV}{V}\,,
\end{equation}
where we used Eq.~\eqref{eq:epsHid} together with $dN=-\,H\,dt=-\,d\phi/(\dot\phi/H)$ and the change of variables $d\phi=dV/V_{,\phi}$. We adopt the remaining–e–folds convention $N(t)=\int_t^{t_e} H\,dt$, so $N$ decreases to $0$ at $t_e$.
 Here and below, the subscript $e$ denotes evaluation at the end of inflation.

For concreteness we take a monomial potential 
\begin{equation}
V(\phi)=A\,\phi^{n},
\end{equation}
with $n$ a rational number and $A$ a normalization fixed by the scalar power spectrum at the CMB pivot $k_\ast=0.05\,{\rm Mpc}^{-1}$, $\mathcal{P}_\ast\simeq2.1\times10^{-9}$. We denote by $N_\ast$ the value of the remaining e-folds $N$ when the pivot mode exits the sound-horizon, i.e., at $c_s k_\ast=aH$. Using~\eqref{phidoth1} with $C_\rho\equiv(6^{\rho-1}/\rho)^{1/(2\rho-1)}$ and $c_s^2=1/(2\rho-1)$, Eq.~\eqref{e-fold} can be integrated in closed form:
\begin{widetext}
\begin{eqnarray}\label{Nfunction}
\nonumber
N&=& \frac{ c_{s}^{2}\left(\frac{V}{A}\right)^{\frac{\rho c_{s}^{2}}{n}} V^{c_{s}^{2}(\rho-1)}}
{n^{c_{s}^{2}}\, C_\rho\, \mathcal{I}_{11}(\rho,n)\, \mathcal{I}_{32}(\rho,n)}
\bigg[
\frac{\mathcal{I}_{32}(\rho,n)}{c_{s}^{2}}\,F_{1}\!\left(\frac{\mathcal{I}_{11}(\rho, n)}{n},-c_{s}^{2}(1+\rho),2 \rho c_{s}^{2};\frac{\mathcal{I}_{32}(\rho,n)}{n};-2 \mathcal{K} V,-4 \mathcal{K} V \right)
\\[3pt]
&&\hspace{4.0cm}
+\,\frac{4\,\mathcal{K} V \,\mathcal{I}_{11}(\rho,n)}{ c_{s}^{2}}\,
F_{1}\!\left(\frac{\mathcal{I}_{32}(\rho,n)}{n},-c_{s}^{2}(1+\rho),2 \rho c_{s}^{2};\frac{\mathcal{I}_{53}(\rho,n)}{n};-2 \mathcal{K} V,-4 \mathcal{K} V\right)
\bigg],
\end{eqnarray}
\end{widetext}
where $\mathcal{I}_{ab}(\rho,n)\equiv\big(n(a\rho-b)+2\rho\big)/(2\rho-1)$ and $F_{1}$ is the Appell hypergeometric function~\cite{whittaker1990course}. The $V$ appearing in~\eqref{Nfunction} is evaluated at sound-horizon exit (i.e., at $N_\ast$).

The complexity of $F_1$ prevents an exact inversion $V(N)$ in closed form. However, in the small-mixing regime $|\mathcal{K}|V\ll1$ (enforced by $f>0$), one can expand and invert perturbatively. Writing $\epsilon \equiv A^{\frac{2 \rho c_{s}^{2}}{\mathcal{I}_{11}}}\,\mathcal{K}$, $\kappa(\rho,n,N)\equiv \Big(C_\rho\,N\,n^{c_{s}^{2}}\mathcal{I}_{11}\Big)^{\!\frac{n}{\mathcal{I}_{11}}}$, one finds
\begin{eqnarray}\label{Vfunction}
\begin{aligned}
V \approx \kappa\,A^{\frac{2 \rho c_{s}^{2}}{\mathcal{I}_{11}}}
\Bigg[1
-\frac{2 c_{s}^{2} n (\rho-1) }{\mathcal{I}_{32}} \,\kappa\, \epsilon
\\
+ \frac{2 c_{s}^{6} n}{{\mathcal{I}_{32}}^{2} \mathcal{I}_{53}}
\Big(4 \rho^{2}(\rho-3) + 2 n \rho \big(15+ 9 \rho^{2}-28\rho\big)\\
+\,n^{2} \big(\rho\,(69+22 \rho^{2}-73\rho)-20\big)\Big)
(\kappa \epsilon)^{2}
+\mathcal{O}(\epsilon^{3})\Bigg].
\end{aligned}
\end{eqnarray}

Combining~\eqref{epsilonv} with~\eqref{Vfunction} gives the slow-roll parameters directly as functions of $N$. To second order in $\epsilon$ one obtains
\begin{widetext}
\begin{eqnarray}\label{epsilonNew}
\nonumber
\varepsilon_H &\approx& \frac{n}{2\,\mathcal{I}_{11}\,N}
\Bigg[1+\frac{2 c_{s}^{4}}{\mathcal{I}_{32}}\,\mathcal{I}_{21}\,\kappa\,\epsilon
- \frac{4 c_{s}^{6}}{ {\mathcal{I}_{32}}^{2} \mathcal{I}_{53}}
\Big(8 \rho^{3}+24 n \rho^{2} (2 \rho-1)+2 n^{2} \rho \big(12+\rho(47 \rho-50)\big)
\\
&&\hspace{6.7cm}
+\,n^{3} \big(\rho(48+ 58 \rho^{2}-95\rho)-7\big)\Big) \kappa^{2} \epsilon^{2}
+\mathcal{O}(\epsilon^{3})\Bigg],\\[4pt]
\nonumber
\eta_H&=&-\frac{1}{\varepsilon_H}\frac{d\varepsilon_H}{d N}
\approx \frac{1}{N}\Bigg[1-\frac{2 n c_{s}^{4}}{\mathcal{I}_{32}\mathcal{I}_{11}}\,\mathcal{I}_{21}\,\kappa\,\epsilon
+\frac{4 n c_{s}^{6} }{{\mathcal{I}_{32}}^{2} {\mathcal{I}_{53}} \mathcal{I}_{11}}
\Big(n^{2} \rho(62+113n)-17 n^{3}
- 2 n \rho^{2} (34+3n (42+37 n))
\\
&&\hspace{7.4cm}
+\,4 \rho^{3} (1+2n) (6+n (21+17 n))\Big)\kappa^{2} \epsilon^{2}
+\mathcal{O}(\epsilon^{3})\Bigg].
\end{eqnarray}
\end{widetext}

For $k$–essence the scalar and tensor spectra at sound-horizon exit ($c_s k=aH$) read
\begin{equation}\label{powerR}
\mathcal{P}_{\mathcal{R}}\simeq \left.\frac{H^{2}}{8\pi^{2}\,\varepsilon_{H}\,c_{s}}\right|_{c_s k=aH},
\qquad
\mathcal{P}_{t}=\left.\frac{2H^{2}}{\pi^{2}}\right|_{k=aH}.
\end{equation}
To first order in slow roll (so that $d\ln k\simeq -\,dN$) it is standard to define
\begin{equation}\label{rr}
\begin{aligned}
r\equiv \frac{\mathcal{P}_{t}}{\mathcal{P}_{\mathcal{R}}}=16\,\varepsilon_H\,c_s,\\
n_s-1\equiv \frac{d\ln \mathcal{P}_{\mathcal{R}}}{d\ln k}
=-\frac{d\ln \mathcal{P}_{\mathcal{R}}}{dN}
= -\,2\varepsilon_H-\eta_H-s,  
\end{aligned}
\end{equation}
with $s\equiv \dot c_s/(Hc_s)=0$ for~\eqref{Lmnon}. Using~\eqref{epsilonNew} yields the analytic estimate
\begin{equation}\label{nsanalytic}
\begin{aligned}
n_s-1 = -\Big(1+\frac{n}{\mathcal{I}_{11}}\Big)\frac{1}{N}
\Bigg[1+\frac{4 n c_{s}^{6} }{{\mathcal{I}_{32}}^{2} {\mathcal{I}_{53}} }
\Big(8 \rho^{2} + 2 n \rho (15 \rho-7)\\
+n^{2}\big(5+ 26 \rho^{2}-25\rho\big)\Big)\kappa^{2} \epsilon^{2}
+\mathcal{O}(\epsilon^{3})\Bigg],
\end{aligned}
\end{equation}
which smoothly reproduces the separable non-canonical result~\cite{Li_2012} in the $\epsilon\to0$ limit. From~\eqref{rr} and~\eqref{epsilonNew}, $r(N)$ follows immediately. For $\mathcal K<0$ (so that $2f-1<1$ for $V>0$), the $\epsilon$–corrections typically reduce both $n_s$ and $r$ relative to the separable case, moving predictions toward the current ACT–preferred region.

\section{Results}\label{Sec3}
In this section we confront the analytic formulae derived in Secs.~\ref{Sec1}--\ref{Sec2} with exact background integrations, and interpret them in light of recent constraints on the inflationary parameters—chiefly $n_s$ and $r$—from ACT in combination with Planck, BK18, and BAO (P--ACT--LB)~\cite{ACT:2025fju,ACT:2025tim}. Throughout, we evaluate observables at sound-horizon exit ($c_s k=aH$). For each potential the normalization $A$ is fixed by matching the scalar amplitude at the CMB pivot, $\mathcal{P}_{\mathcal R}(k_\ast)\simeq 2.1\times10^{-9}$ at $k_\ast=0.05\,\mathrm{Mpc}^{-1}$. Once $A$ is fixed, we scan the small-mixing parameter $\epsilon_{\rm mix}$ (which coincides with $\epsilon$ defined in Sec.~\ref{Sec2}):
\[
\epsilon_{\rm mix}\equiv A^{\frac{2\rho c_s^2}{\mathcal{I}_{11}}}\,\mathcal{K},
\]
which effectively corresponds to scanning the coupling $\mathcal{K}$ at fixed scalar amplitude. The $e$-fold variable $N$ counts the number of $e$-folds remaining to the end of inflation, $N=\int_t^{t_e} H\,dt$, so that $N_\ast\in[50,70]$ brackets the usual reheating uncertainty.

\begin{table}[t!]
  \centering
  \caption{Numerical (subscript $N$; exact background integration) vs.\ analytic (subscript $A$; expansion from Eqs.~\eqref{epsilonNew}, \eqref{rr}, \eqref{nsanalytic}) predictions for $n_s$ and $r$ at sound-horizon exit for two benchmarks: Model I $(\rho=10,\,n=2,\,N_\ast=70)$ and Model II $(\rho=2,\,n=\tfrac{2}{3},\,N_\ast=60)$. The agreement is better than $0.2\%$ for $n_s$ and at the few-percent level for $r$ across the scanned $\epsilon_{\rm mix}$ range.}
  \label{table2}
  \renewcommand{\arraystretch}{1.12}
  \setlength{\tabcolsep}{5pt}
  \begin{ruledtabular}
  \begin{tabular}{l c cc cc}
    & & \multicolumn{2}{c}{Numerical} & \multicolumn{2}{c}{Analytic} \\
    \cline{3-4}\cline{5-6}
    Model & $|\epsilon_{\rm mix}|$ & $n_{s,N}$ & $r_{N}$ & $n_{s,A}$ & $r_{A}$ \\
    \hline
    \multirow{5}{*}{\shortstack[l]{I\\$\rho=10$\\$n=2$\\$N_\ast=70$}}
      & 0.0001 & 0.9714 & 0.0247 & 0.9713 & 0.0248 \\
      & 0.0002 & 0.9707 & 0.0231 & 0.9707 & 0.0232 \\
      & 0.0003 & 0.9691 & 0.0210 & 0.9695 & 0.0211 \\
      & 0.0004 & 0.9659 & 0.0184 & 0.9655 & 0.0184 \\
      & 0.0005 & 0.9575 & 0.0148 & 0.9576 & 0.0149 \\
    \hline
    \multirow{4}{*}{\shortstack[l]{II\\$\rho=2$\\$n=2/3$\\$N_\ast=60$}}
      & 0.010  & 0.9755 & 0.0273 & 0.9755 & 0.0271 \\
      & 0.015  & 0.9739 & 0.0231 & 0.9741 & 0.0230 \\
      & 0.020  & 0.9701 & 0.0179 & 0.9712 & 0.0177 \\
      & 0.025  & 0.9578 & 0.0110 & 0.9577 & 0.0107 \\
  \end{tabular}
  \end{ruledtabular}
\end{table}

\begin{figure}[t]
\centering
\includegraphics[width=0.45\textwidth]{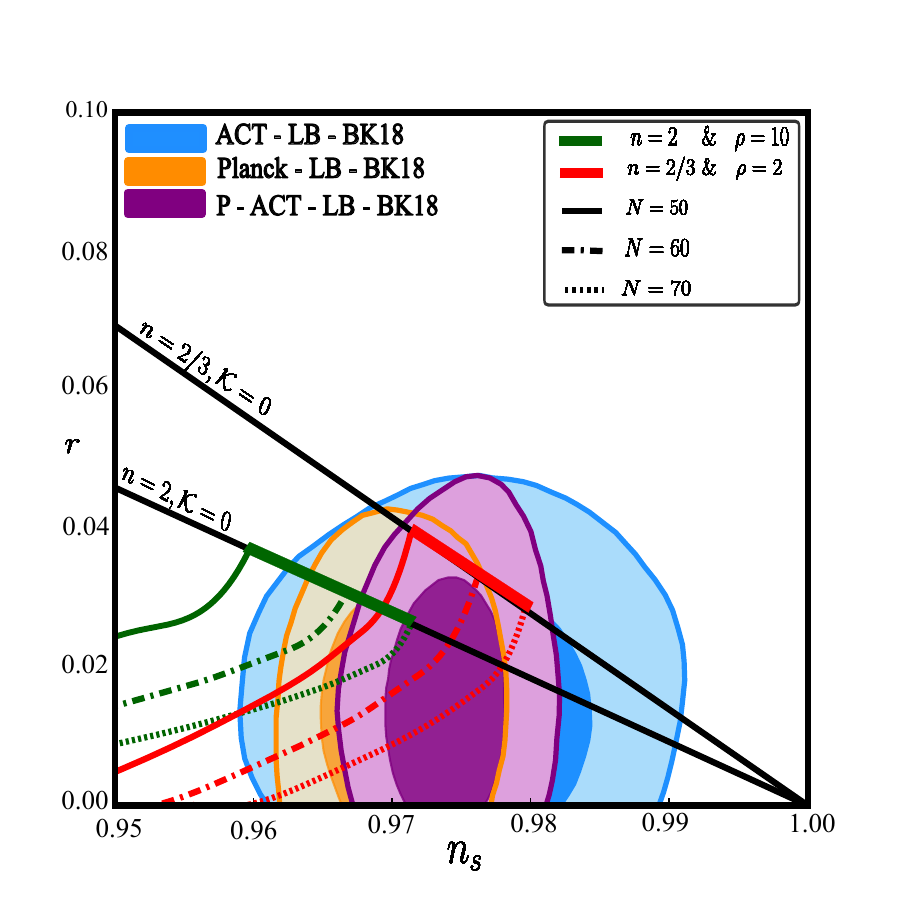}
\caption{Predictions in the $(n_s,r)$ plane from Eqs.~\eqref{epsilonNew}, \eqref{rr} and~\eqref{nsanalytic}. Curves correspond to varying $\epsilon_{\rm mix}$ at fixed $(\rho,n,N_\ast)$; contours combine ACT with Planck 2018, BK18 and BAO (P--ACT--LB).}
\label{fig3}
\end{figure}

We compare the analytic predictions from Eqs.~\eqref{epsilonNew},~\eqref{rr} and~\eqref{nsanalytic} with exact numerical results obtained by solving Eqs.~\eqref{Hnonold} and~\eqref{rhoback} for $\{\phi(N),H(N)\}$.\footnote{In the numerics $N$ decreases to zero at the end of inflation. For each parameter point we verify $f(\phi)>0$ and $\rho>1/2$ (so $c_s^2=1/(2\rho-1)>0$), and we enforce the positivity bound $|\mathcal{K}|V<\tfrac12$, equivalent to $f=1+2\mathcal{K}V>0$.
} We denote the mixing expansion parameter by $\epsilon_{\rm mix}$ to avoid confusion with the Hubble slow-roll parameter $\varepsilon_H$.

The agreement in Table~\ref{table2} is excellent: typical (maximal) fractional differences are $\lesssim 3\times10^{-4}$ ($\sim 1.1\times10^{-3}$) for $n_s$ and $\sim 0.8\%$ ($\sim 2.7\%$) for $r$, validating the $\mathcal{O}(\epsilon_{\rm mix}^2)$ expansion. For fixed $(\rho,n,N_\ast)$, dialing $\mathcal{K}<0$—and thus $2f-1<1$ for $V>0$—systematically shifts both $n_s$ and $r$ downward relative to the separable limit, as anticipated analytically. On the data side, the ACT+Planck+BAO combination (P--ACT--LB) prefers $n_s=0.9743\pm0.0034$ with $r<0.038$ (95\% CL)~\cite{ACT:2025fju,ACT:2025tim}; several entries in Table~\ref{table2} fall squarely in this window. The model flow in the $(n_s,r)$ plane obtained from Eqs.~\eqref{epsilonNew} and~\eqref{nsanalytic}, with observational contours overlaid, is shown in Fig.~\ref{fig3}.

\begin{figure*}[t]
\centering
\includegraphics[width=0.45\textwidth]{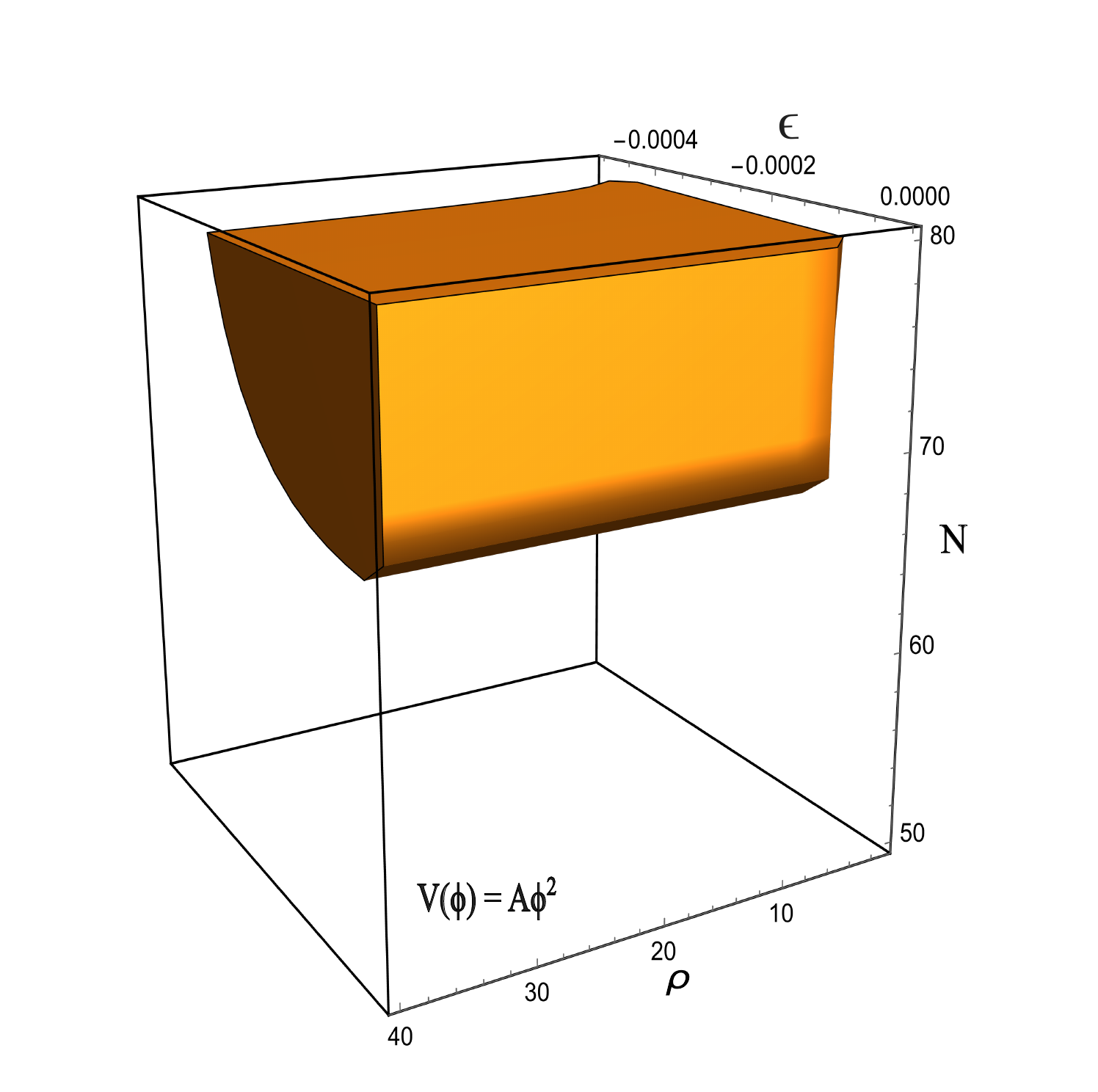}\vspace{4pt}
\includegraphics[width=0.45\textwidth]{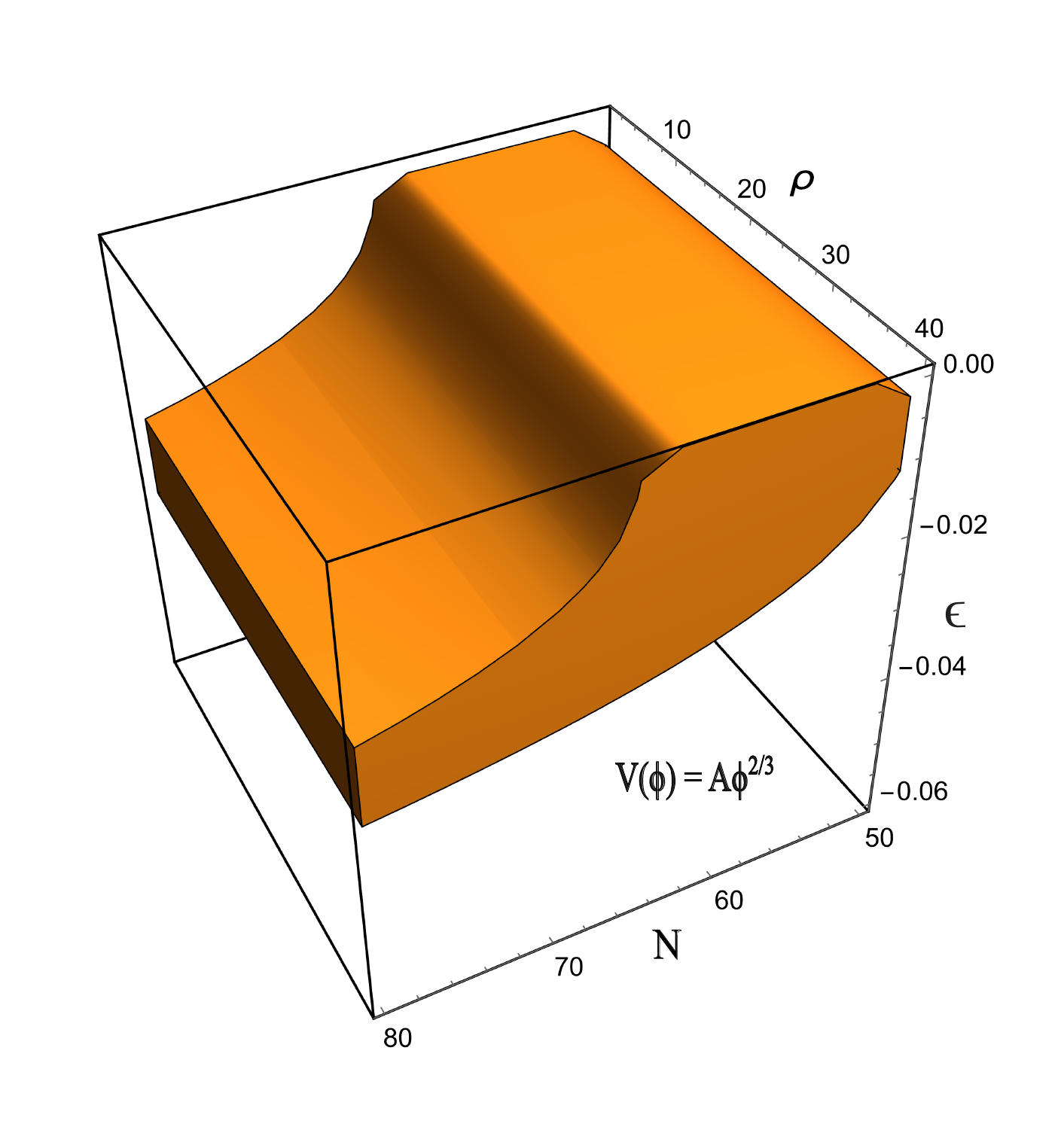}
\caption{Allowed region in $(\epsilon_{\rm mix},\rho,N_\ast)$ for $n=2$ (top) and $n=2/3$ (bottom). Shaded points satisfy the P--ACT--LB constraints on $(n_s,r)$. All shown points also obey $f>0$ and $|\mathcal{K}|V<\tfrac12$.}
\label{fig1}
\end{figure*}

\begin{figure*}[t]
\centering
\includegraphics[width=0.45\textwidth]{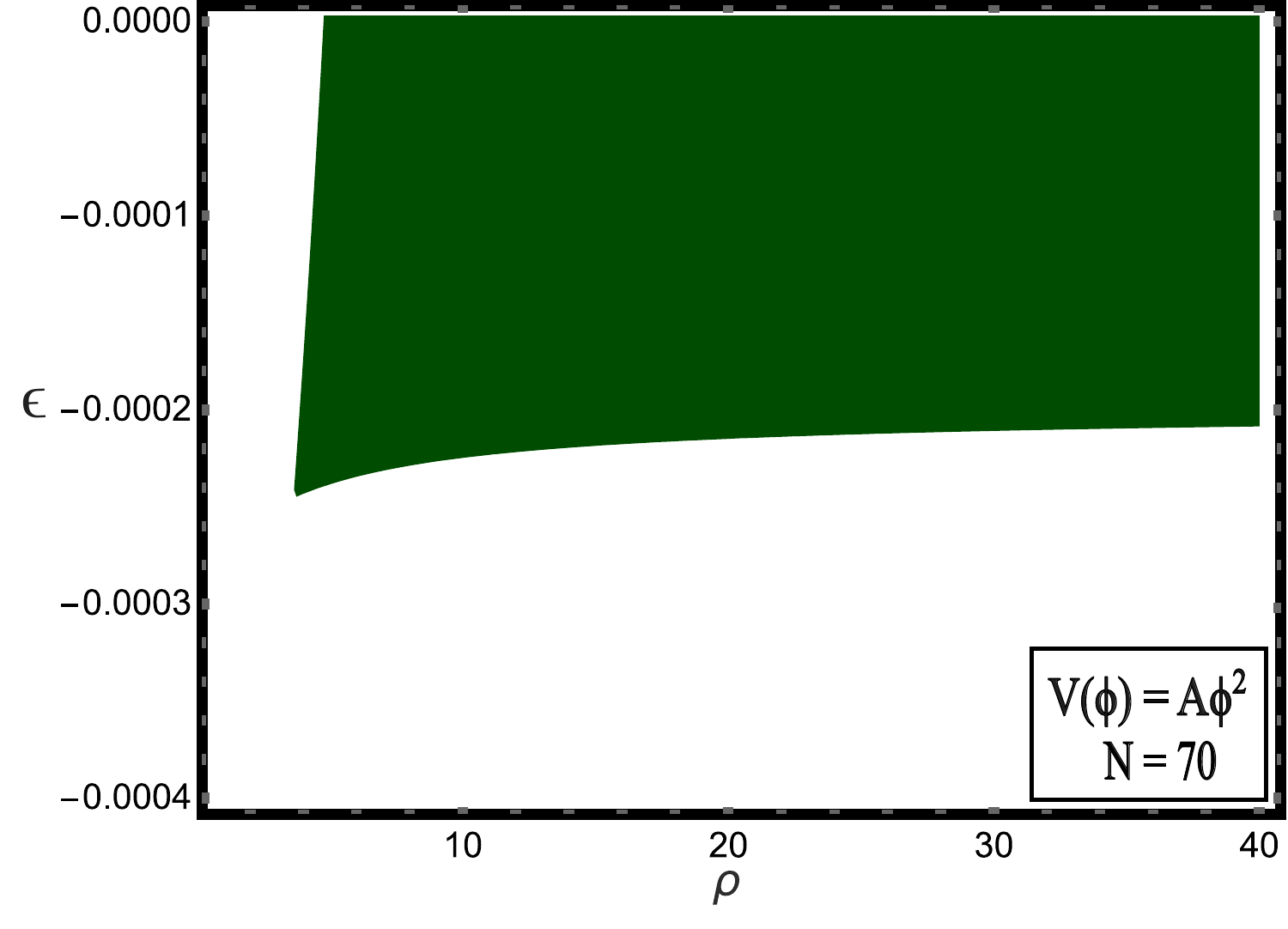}\vspace{4pt}
\includegraphics[width=0.45\textwidth]{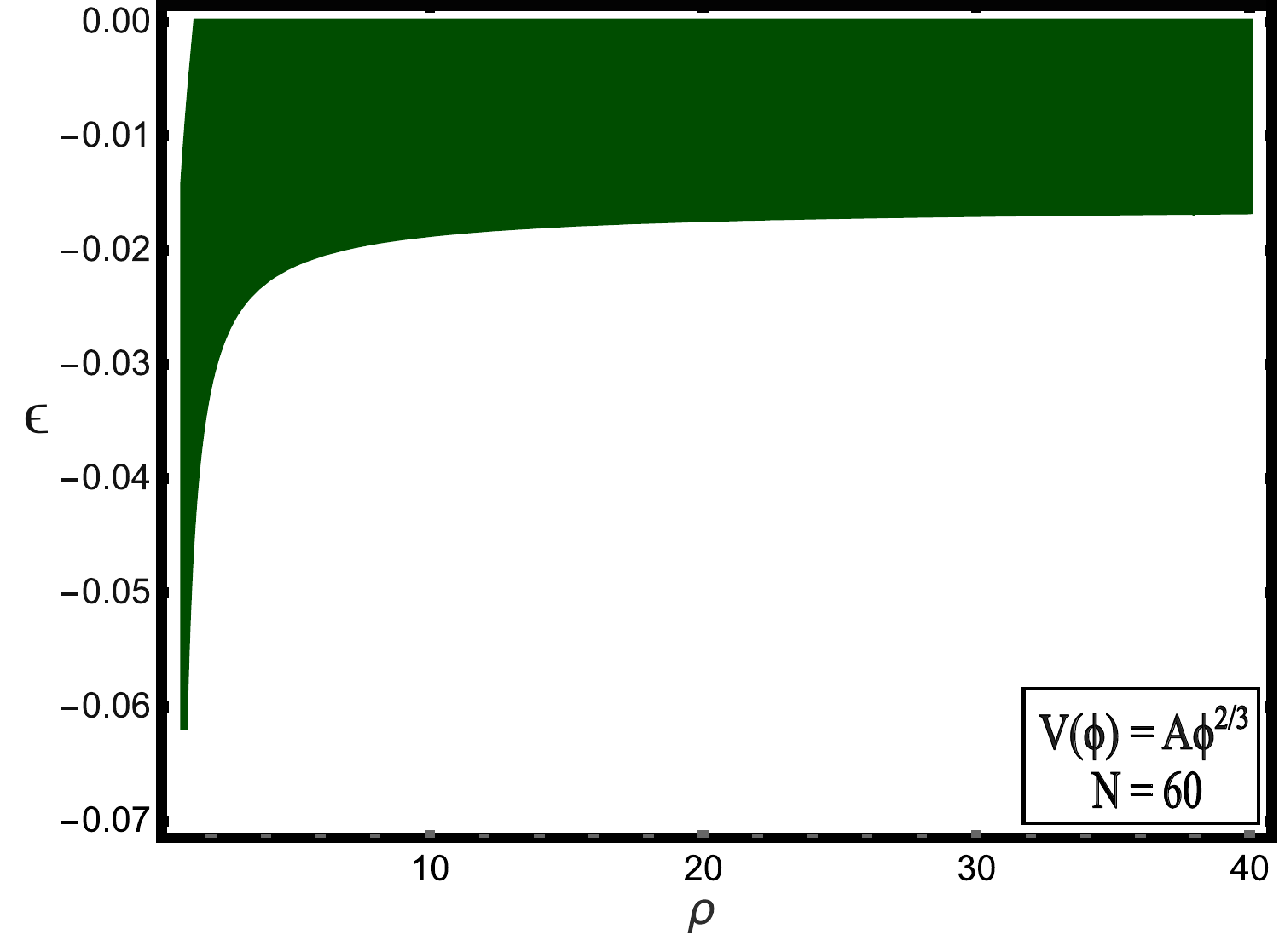}
\caption{Slices of Fig.~\ref{fig1} showing the allowed $(\epsilon_{\rm mix},\rho)$ region for fixed $N_\ast$ ($N_\ast=70$ for $n=2$ and $N_\ast=60$ for $n=2/3$). Shaded areas lie inside the P--ACT--LB contour.}
\label{fig2}
\end{figure*}

A broader view of parameter dependence is provided in Fig.~\ref{fig1}, which displays the allowed region in $(\epsilon_{\rm mix},\rho,N_\ast)$ for $n=2$ and $n=2/3$. Increasing $\rho$ (thereby decreasing $c_s$) and/or increasing $|\mathcal{K}|$ lowers $r$ at fixed $N_\ast$, while $n_s$ is chiefly controlled by $N_\ast$ and only gently shifted by $\epsilon_{\rm mix}$. In particular, within the scans shown we find that agreement with P--ACT--LB typically requires
\[
\begin{aligned}
\text{for } V=A\phi^2:\;& N_\ast\gtrsim 70,\quad \rho\gtrsim 10,\\[-2pt]
& -6.2\times10^{-4}<\epsilon_{\rm mix}\le 0,\\[4pt]
\text{for } V=A\phi^{2/3}:\;& N_\ast\gtrsim 50,\quad \rho\gtrsim 1,\\[-2pt]
& -3.19\times10^{-2}<\epsilon_{\rm mix}\le 0~.
\end{aligned}
\]
Two-dimensional slices at fixed $N_\ast$ are shown in Fig.~\ref{fig2}.

Finally, we assess robustness within the same narrative flow. Varying $N_\ast$ in the canonical range $[50,70]$ shifts $n_s$ roughly as $\Delta n_s\simeq (1+n/\mathcal{I}_{11})\,\Delta N_\ast/N_\ast^2$, while $r$ scales as $1/N_\ast$ at leading order; the qualitative impact of the non-separable mixing is unchanged across this interval. Because $c_s^2=1/(2\rho-1)$ is constant in our class, the equilateral-type non-Gaussianity scales as $f_{\rm NL}^{\rm equil}\sim\mathcal{O}(c_s^{-2}-1)$ with an $\mathcal{O}(1)$ coefficient set by the cubic couplings of $P(X,\phi)$~\cite{Chen_2007}. For the benchmarks displayed ($\rho=2$ and $\rho=10$) this corresponds to amplitudes $\mathcal{O}(1\text{--}10)$, comfortably consistent with current bounds. All parameter points in the figures and table satisfy $f(\phi)>0$ and $\rho>1/2$ (no ghosts, $c_s^2>0$), and respect the hard upper bound $V<1/(2|\mathcal{K}|)$ derived in Sec.~\ref{Sec1}. In short, the non-separable mixing controlled by $\mathcal{K}<0$ systematically lowers both $n_s$ and $r$ relative to the separable non-canonical limit for the monomials tested, enabling otherwise marginal potentials to fit the P--ACT--LB contours while remaining theoretically healthy in the slow-roll regime.

\section{Conclusion}\label{Sec4}

We studied a non-separable subset of $k$-essence in which the kinetic and potential sectors interact through an $X^{\rho}V(\phi)$ coupling implemented via a potential-dependent prefactor $f(\phi)=1+2\mathcal K V$. In slow roll this structure modifies the Hubble flow in a controlled way and shifts the inflationary predictions relative to the separable non-canonical template. For monomial potentials $V=A\phi^n$ (with $n=2$ and $n=2/3$ as representative cases) we derived closed analytic expressions for $n_s(N_\ast)$ and $r(N_\ast)$ to $\mathcal{O}(\epsilon_{\rm mix}^2)$, with $\epsilon_{\rm mix}\propto\mathcal K$, and verified them against exact background integrations. The analytic and numerical predictions agree at the sub-per-mille level for $n_s$ and at the percent level for $r$, validating the small-mixing expansion. Throughout, the class has constant sound speed $c_s^2=1/(2\rho-1)$ and satisfies the standard health conditions provided $f(\phi)>0$, $\rho>\tfrac12$, and $V<1/(2|\mathcal K|)$ (the positivity bound from $f>0$).

A robust qualitative outcome is that for $\mathcal K<0$ (hence $2f-1<1$ along the trajectory) the non-separable mixing systematically lowers both $n_s$ and $r$ at fixed $N_\ast$, moving otherwise marginal monomials toward the region favored by recent ACT+Planck+BK18+BAO constraints (P--ACT--LB). Our parameter scans illustrate viable domains without performing a likelihood fit, and they clarify the roles of $N_\ast$ (primarily setting $n_s$), $\rho$ (via $c_s$), and the mixing strength $\epsilon_{\rm mix}$ (tuning both observables). Because $c_s$ is constant, the leading equilateral bispectrum scales as $\mathcal{O}(c_s^{-2}-1)$ with an $\mathcal{O}(1)$ coefficient set by the cubic couplings of $P(X,\phi)$; for the benchmarks considered, this implies amplitudes comfortably consistent with current bounds. 

A natural continuation is a full statistical confrontation with the P--ACT--LB likelihood (and upcoming CMB-S4/Simons Observatory data), a dedicated bispectrum analysis, and an exploration of broader potentials (e.g., $V=V_0 e^{-\alpha \phi^n}$, $n>1$) and UV completions that generate the $X^{\rho}V(\phi)$ interaction. Possible connections to late-Universe tensions (e.g., $H_0$, $S_8$) lie outside the scope of the present inflationary study but motivate complementary work in related $XV$ scenarios.

\acknowledgments
\"{O}.A.\ acknowledges the support of the Turkish Academy of Sciences in the scheme of the Outstanding Young Scientist Award (T\"{U}BA-GEB\.{I}P). MS is supported by the Science and Engineering Research Board (SERB), DST, Government of India under the Grant Agreement number CRG/2022/004120 (Core Research Grant). MS is also partially supported by the Ministry of Education and Science of the Republic of Kazakhstan, Grant No. 0118RK00935.

\bibliography{inflation-ACT}
	
\end{document}